\documentclass{emulateapj}
\usepackage{amsmath}
\usepackage{xspace}

\newcommand{\rbr}[1]{\ensuremath{\left( #1 \right)} }

\newcommand{\abs}[1]{\ensuremath{\left| #1 \right|} }
\newcommand{\E}[1]{\ensuremath{\times 10^{#1}} }

\newcommand{\Tasc}{\ensuremath{T_{\text{asc}}}}
\newcommand{\ob}{\ensuremath{\mathcal{O}}}

\newcommand{\conf}{\ensuremath{\mathcal{C}}}

\newcommand{\Msol}{ \ensuremath{ M_{\odot} } }
\newcommand{\sps}[1]{s~s$^{-#1}$}
\newcommand{\hps}[1]{Hz~s$^{-#1}$}

\usepackage{xcolor}
\slugcomment{Draft Version \today.}

\begin{document}

\title{Coherent timing of the accreting millisecond pulsar NGC 6440 X-2}

\author{Peter Bult\altaffilmark{1}}
\author{Alessandro Patruno\altaffilmark{2,3}}
\author{Michiel van der Klis\altaffilmark{1}}

\altaffiltext{1}{	
	Anton Pannekoek Institute, University of Amsterdam,
	Postbus 94249, 1090 GE Amsterdam, The Netherlands
}
\altaffiltext{2}{
	Leiden Observatory, Leiden University,
	Postbus 9513, 2300 RA Leiden, The Netherlands
}
\altaffiltext{3}{
	ASTRON, the Netherlands Institute for Radio Astronomy, 
	Postbus 2, 7900 AA, Dwingeloo, The Netherlands
}

%% Abstract
\begin{abstract}
We study the 205.9 Hz pulsations of the accreting millisecond X-ray pulsar NGC 6440 X-2 across all outbursts observed with the {\it Rossi X-ray Timing Explorer} over a period of 800 days. We find the pulsations are highly sinusoidal with a fundamental amplitude of $5\%-15\%$ rms and a second harmonic that is only occasionally detected with amplitudes of $\lesssim2\%$ rms.  By connecting the orbital phase across multiple outbursts, we obtain an accurate orbital ephemeris for this source and constrain its 57 min orbital period to sub-ms precision. We do not detect an orbital period derivative to an upper limit of $|\dot{P}| \leq 8 \times 10^{-11}~{\rm s~s}^{-1}$.  We investigate the possibility of coherently connecting the pulse phase across all observed outbursts, but find that due to the {poorly constrained} systematic uncertainties introduced by a flux-dependent bias in the pulse phase, multiple statistically acceptable phase-connected timing solutions exist.
\end{abstract}

%% Keywords 

%%--------------------------------------------------------------------------------------------
%%                MAIN BODY
%%--------------------------------------------------------------------------------------------
\section{Introduction} 
    Accreting Millisecond X-ray Pulsars (AMXPs) are rapidly rotating neutron
    stars in low-mass X-ray binaries. These systems show coherent X-ray
    pulsations that arise when the accretion flow is magnetically channeled 
    to the stellar surface. These pulsations offer a physical tracer
    of the neutron star and the inner accretion flow toward it, as their
    frequency gives a direct measure of the neutron star rotation rate, and the
    shape of the pulse waveform carries information on the accretion geometry
    and compactness of the neutron star \citep{Poutanen2003, Leahy2008, 
    Psaltis2014a}. Tracking the pulse arrival times allows to measure the
    evolution of the neutron star spin and the binary orbit, thus offering
    insight into accretion torque theory \citep{Psaltis1999}, alternative
    torque mechanisms \citep{Bildsten1998b, Haskell2011} and the binary evolution
    of millisecond pulsars \citep{Bildsten2002, Nelson2003, Patruno2012a}.
	
    Among the currently known AMXPs (see \citealt{Patruno2012b} for a review), the
    globular cluster source NGC 6440 X-2 is unique in its outburst
    behavior; it shows comparatively short, low luminosity outbursts, with peak
    X-ray luminosities of $L_X \lesssim 1.5\E{36}$~erg~s$^{-1}$ and outburst
    durations of $2-5$~days \citep{Heinke2010}.  NGC 6440 X-2 was discovered
    with Chandra on July~28th, 2009 \citep{Heinke2009} and seen in outburst 
    again merely a month later with the {\it Rossi X-ray Timing Explorer} ({\it RXTE}), at which 
    time the $205.9$~Hz pulsations were discovered \citep{Altamirano2009}. The 
    following two outbursts each occurred after a quiescent interval of about 
    one month, establishing NGC 6440 X-2's
    recurrence time as the shortest of all AMXPs known to date. The coherent
    timing analysis of those first four outbursts was reported by
    \citet{Altamirano2010a}, who found pulsations in three outbursts at
    fractional amplitudes of $\sim7$\% for the fundamental component.
	
    After the fourth outburst, on October 28th, 2009, {no activity from 
    NGC 6440 X-2 was observed until} March, 2010, {although outbursts
    may have been missed due to visibility constraints and activity from
    other X-ray sources the same field} \citep{Altamirano2010b}.
	Subsequently, the source showed another three
    outbursts with a recurrence time of about 110 days, after which it remained
    {undetected} for nearly 300~days until the last outburst
    was observed with {\it RXTE} in November, 2011 \citep{Patruno2013}.
    
    In this work we present the results of a coherent timing analysis of NGC~6440~X-2 over the
    course of its complete outburst history as observed with {\it RXTE}. In
    Section~\ref{sec:datareduction} we describe our data reduction and analysis
    method. In Section~\ref{sec:results} we present our results as we discuss
    how the high precision orbital ephemeris and pulse frequency evolution of
    NGC~6440~X-2 was obtained. Next, in Section~\ref{sec:discussion}, we
    briefly summarize and discuss our results.

\section{Data Reduction}
\label{sec:datareduction}
    We analyze all pointed {\it RXTE} observations of NGC 6440 X-2 \citep{Altamirano2009,
    Heinke2010, Patruno2013}. We use the 16-s time-resolution Standard-2 data to construct 
    2--16 keV light curves, {normalized to the count rate of the Crab {($\sim2300$ 
    ct s$^{-1}$ PCU$^{-1}$; $\sim2.8\E{-8}$ erg cm$^{-2}$ s$^{-1}$)}} and averaged per 
    observation (see, e.g, \citealt{Straaten2003} for details). 
    We find one type I X-ray burst at MJD 55359.5, which we exclude from our further analysis.
    
\begin{deluxetable*}{lllrlcl}
	% Preamble
    \newcommand{\ph}{\phantom{1}}
    \tabletypesize{\scriptsize}
    \tablecaption{ Outbursts of NGC 6440 X-2
    \label{tab:outbursts} }
    \tablewidth{1.0\linewidth}
    
    % Header
    \tablehead{%
		ID & Date & \Tasc & $\Delta$MJD & ObsID & Exposure & Pulse Amp. \\
		~  & ~    & (MJD) & (days)      & ~     & (s)      & (\% rms)
	}
	
	% Data
	\startdata
		\ob2  & 2009-07-28 & $55042.81$  &       & 94044-04-01-00$^\text{a}$  & \phantom{1} 1900  & $12.9\pm 4.1$ \\ 
		\ob3  & 2009-08-30 & $55073.03$  &  30.3 & 94044-04-02-00$^\text{b}$  & \phantom{1} 3200  & $9.0 \pm 0.5$ \\ 
              & \nodata    & \nodata     &       & 94044-04-02-01             & \phantom{} 14000  & $5^\text{c}$ \\
        \ob4  & 2009-10-01 & $55106.01$  &  32.9 & 94044-04-03-00             & \phantom{1} 2200  & $11.4 \pm 1.9$ \\
              & \nodata    & \nodata     &       & 94044-04-04-00             & \phantom{1} 3400  & $16^\text{c}$ \\
        \ob5  & 2009-10-28 & $55132.90$  &  27.0 & 94315-01-04-01             & \phantom{14} 900  & $14.0 \pm 1.8$ \\
              & \nodata    & \nodata     &       & 94315-01-04-02             & \phantom{14} 900  & $10^\text{c}$         \\
        \ob6  & 2010-03-21 & $55276.62$  & 143.3 & 94315-01-12-00             & \phantom{1} 2000  & $12.4 \pm 0.7$ \\
              & \nodata    & \nodata     &       & 94315-01-12-01             & \phantom{1} 2700  & $13.2 \pm 1.0$ \\
              & \nodata    & \nodata     &       & 94315-01-12-02             & \phantom{1} 1900  & $12.9 \pm 1.1$ \\
        \ob7  & 2010-06-12 & $55359.47$  &  83.2 & 94315-01-14-00             & \phantom{1} 9900  & $ 7.6 \pm 0.4$ \\
        \ob8  & 2010-10-04 & $55473.85$  & 114.4 & 94315-01-25-00             & \phantom{1} 1200  & $ 5.3 \pm 0.6$ \\
        \ob9  & 2011-01-23 & $55584.71$  & 110.9 & 96326-01-02-00             & \phantom{1} 2100  & $ 6.0 \pm 0.6$ \\
        \ob10 & 2011-03-21 & $55641.02$  &  56.3 & 96326-01-10-00             & \phantom{1} 2000  & $12.7 \pm 1.9$ \\
        \ob11 & 2011-11-06 & $55871.23$  & 231.9 & 96326-01-35-00             & \phantom{1} 1200  & $ 6.3 \pm 1.1$ \\ 
              & \nodata    & $55872.83$  &       & 96326-01-40-00             & \phantom{1} 3300  & $11.5 \pm 2.3$ \\ 
              & \nodata    & $55873.31$  &       & 96326-01-40-01             & \phantom{1} 2400  & $ 9.8 \pm 2.9$
	\enddata
	
	% Postamble
    \tablenotetext{a}{GoodXenon} \tablenotetext{b}{SingleBit} \tablenotetext{c}{95\% c.l. upper limit}
    \tablecomments{%
    	Pulse amplitudes are that of the fundamental component and calculated per {\it RXTE}
    	orbit ($\sim3$ ks intervals). $\Delta\mbox{MJD}$ gives the duration of the quiescent interval
		with respect to the last observed outburst. Reported ObsIDs refer to Event mode data unless otherwise
		specified. 
    }
\end{deluxetable*}

    For the timing analysis we use all high time resolution ($\leq122\mu$s) SingleBit, 
    Event and GoodXenon data. We correct the data to the Solar System barycenter using
    the \textsc{ftool} \textit{faxbary} based on the Chandra X-ray position of
    \citet{Heinke2010}. This tool also applies the {\it RXTE} fine clock corrections,
    allowing for an absolute timing precision of $\sim\nobreak4~\mu$s \citep{Rots2004}. 
    
    The observations are divided in $\sim500$ s segments, selecting only the events in 
    the energy channels 5--37 ($\sim2-16$ keV), which optimizes the pulse signal to noise
    ratio. 
    We then compare the observed count rate with the background count rate estimated using 
    the \textsc{ftool} \textit{pcabackest}, and reject all observations for which the expected 
    amplitude of a 100\% modulated source contribution cannot be detected above the
    noise amplitude expected from counting statistics.
    
    The remaining observations are corrected for the orbital ephemeris, folded on the pulse 
    period (see Section \ref{sec:results}), and fit with a constant plus a sinusoid at the 
    fundamental ($\nu$) and second harmonic ($2\nu$) pulse frequency. 
	We consider a pulse harmonic to be significant if its amplitude 
    exceeds a detection threshold, which we define as the noise amplitude for which
    there is only a small probability $1-\conf$ that among all observations one or more exceed 
    it by chance, {and given a $\chi^2$ distribution with 2 degrees of freedom for 
    the squared amplitude of the noise}. {For a confidence level of $\conf=99\%$ and
    and 500 second segments we then find a $A/\sigma_A = 3.8$ detection threshold, where
    $A$ is the pulse amplitude and $\sigma_A$ its uncertainty}. Once an episode of pulsations
    has been established we repeat the analysis for several different segment lengths
    ($100-3000$ s) to study the pulse properties on various timescales.
    Pulse amplitudes are reported in terms of fractional rms
    \begin{equation}
		r_i = \frac{1}{\sqrt{2}} \frac{A_i}{N_{\gamma}-B},
	\end{equation}
	where $A_i$ is the measured sinusoidal amplitude of the $i$-th harmonic, 
	$N_\gamma$ the total number of counts in the segment and $B$ the estimated 
	background contribution (see, e.g. \citealt{Patruno2010c}).
    
	We model the measured pulse arrival times per outburst using a circular orbit and constant
	pulse frequency. As such our timing model consists of four parameters, namely the orbital 
	period $P_b$, the projected semi-major axis $A_x \sin(i)$, the time of the ascending node
	\Tasc, and the pulse frequency $\nu$. Phase residuals are obtained by subtracting
	this timing model from the measured arrival times and analysed to refine the timing
	model. The details of this analysis are presented in the following section.

\section{Results} 
\label{sec:results}
	Because of the short duration of its outbursts and its high recurrence rate, the coherent timing
	analysis of NGC 6440 X-2 requires careful consideration. In this section we describe 
	our analysis of this source in detail. We start from a provisional timing solution based
	on the work of \citet{Altamirano2010a}, and iteratively refine the timing model parameters until
	the final timing solution is obtained.

		Initially we consider the outbursts of NGC 6440 X-2 separately (see 
		Table \ref{tab:outbursts} for details). We refer to $i$-th outburst as \ob{$i$}, where
		we start counting {outbursts} from an archival {\it Swift} observation on the 4th of June, 2009 
		(MJD 54986), which is considered the earliest detection of this source
		\citep{Heinke2010}. For our preliminary timing solution we adopt the 
		pulse frequency, projected semi-major axis and orbital period that 
		\citet{Altamirano2010a} obtained for \ob3. Because of the limited accuracy 
		of their orbital period measurement ($P_b = 3438(33)$ s) we cannot reliably predict the time of 
		ascending node for any of the other outbursts. Instead, for 
		each outburst, we set this parameter by searching a grid of 
		$\Tasc$ at a 5-second resolution and selecting the value that 
		maximizes the pulse amplitude.

		\begin{figure}[t]
			\centering
			\includegraphics[width=\linewidth]{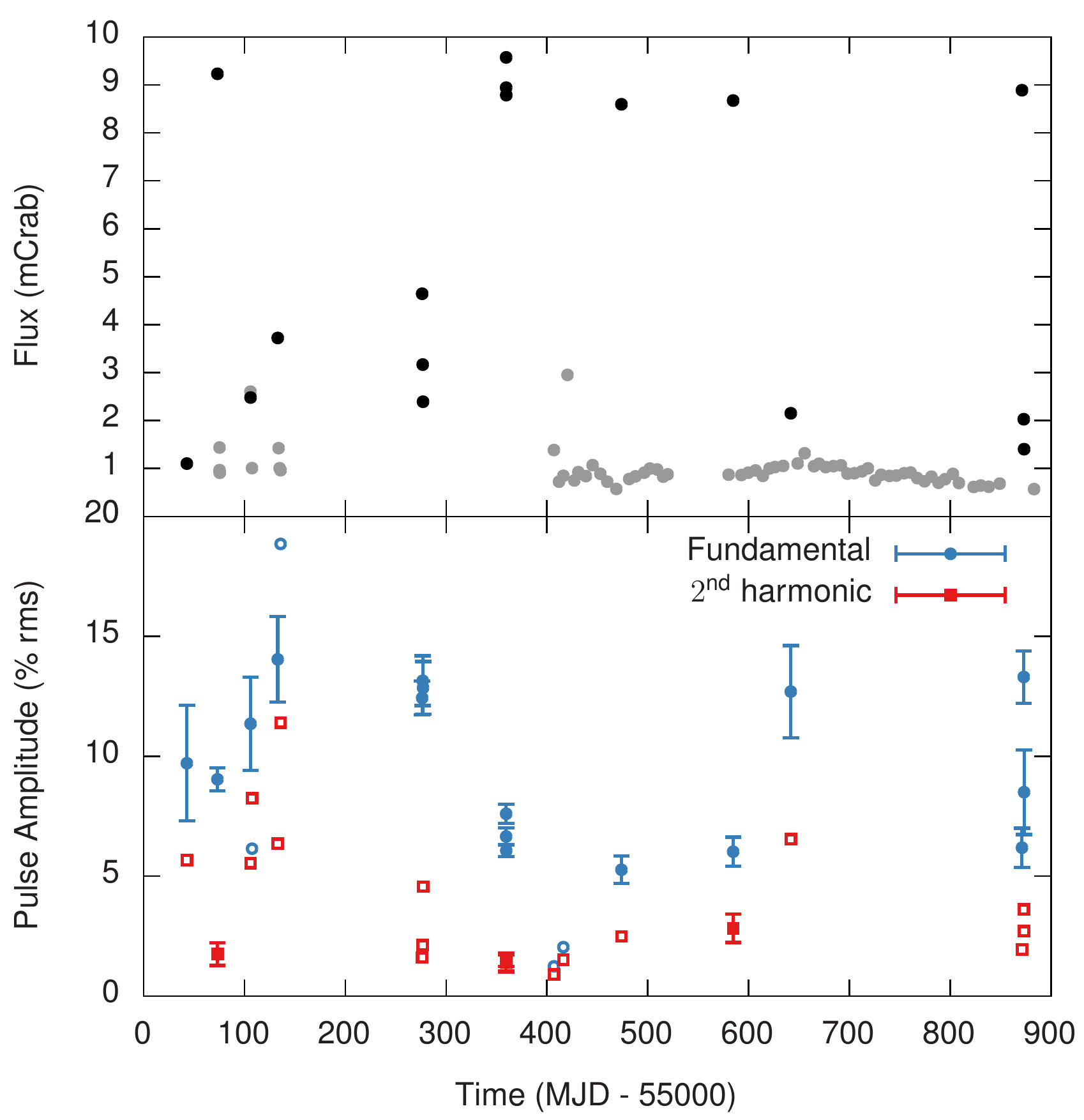}
			\caption{%
				Top: light curve of NGC 6440 X-2 as observed with {\it RXTE}. 
				Black points show observations with significantly
				detected pulsations, gray points show observations for which pulsations
				were not detected.
				Bottom: fundamental (blue circles) and second harmonic (red squares) 
				fractional pulse amplitudes. Open symbols give 95\% upper limits. \\
			}			
			\label{fig:lightcurve}
		\end{figure}
		
		NGC 6440 X-2 has been reported to have shown nine outbursts observed
		with {\it RXTE} during the 800 day observational baseline \citep{Patruno2013}. 
		We find pulsations during this
		baseline on ten occasions, adding one low-flux outburst (\ob10) 
		to the sample.
		Five of the observed outbursts reached a flux of $\sim9$ mCrab, and the other five 
		peaked at fluxes half that or lower (see Figure \ref{fig:lightcurve}). 
		The pulse profiles are highly sinusoidal, with typical fundamental pulse 
		amplitudes of $5\%-15\%$ rms and second harmonic amplitudes of $\sim2\%$ 
		detected for only three observations (\ob3, \ob7, \ob9). We report the 
		details of each outburst in Table \ref{tab:outbursts}.
		
		Although NGC 6440 X-2 has shown many outbursts, only a few observations
		offer sufficient quality to place useful constraints on the timing model 
		parameters. Many of the observations have a low signal to noise ratio
		due to the low count rate of the source. Additionally, most outbursts are 
		observed only for a single exposure of less than 3 ks, which is shorter 
		than the orbital period of the binary. So, even if the phase could be measured
		on a sufficiently short timescale to allow for a fit to our 4-parameter
		timing model, the short exposures introduce strong correlations between
		$\nu$, $T_{\rm asc}$ and $P_b$ and hence cause a large uncertainty on
		all parameters.
		
		For three outbursts of NGC 6440 X-2 the data is of sufficient quality
		to constrain the timing model. These are \ob3, \ob6 and \ob7. The latter
		two are of particular use, as they consist of several consecutive
		exposures with separations of less than half a day. For these outbursts
		the data spans much more than a single orbital period, which breaks
		the correlation between the timing model parameters. We give the
        timing solutions for these outbursts in Table \ref{tab:peroutburst}.
		While \ob11 also consists of multiple exposures, we note that those
		observations are each about a day apart, and that, within the accuracy
		of the frequency measurement, the pulse phase cannot be uniquely
		propagated across that separation.
		
\begin{deluxetable*}{llllll}
    % Preamble
    \tabletypesize{\scriptsize}
    \tablecaption{Per-outburst Timing Solutions
    \label{tab:peroutburst} }
    \tablewidth{1.0\linewidth}

    % Header
    \tablehead{%
        ID &  $\nu$     & $A_x \sin(i)$ & $P_b$  & $T_{\rm asc}$ & $\chi^2/$dof \\[2pt]
           &  (Hz)      & (lt-ms)       & (s) & (MJD)            & 
	}

    % Data
	\startdata
        \ob3	& 205.89217(14) 	& 6.2(6)   & 3421(156)  & 55073.035(3)   & \phantom{1}3/5\phantom{0}   \\[2pt] 	%  3.2  /  5
        \ob6	& 205.8921768(2)	& 6.05(4)  & 3458.0(3)  & 55276.62545(3) & \phantom{1}7/8\phantom{0}   \\[2pt] 	%  6.5  /  8
        \ob7	& 205.8921759(7)    & 6.14(2)  & 3458(1)    & 55359.51080(3) &           14/24   		% 14.2  / 24
    \enddata
\end{deluxetable*}

		\subsection{Orbital evolution}
		To refine the orbital period estimate, we consider the $T_{\rm asc}$ values
		measured locally for \ob3, \ob6 and \ob7 and perform a phase-coherent analysis of the
		orbital evolution.
		Using the timing solution of \ob6 we predict the $T_{\rm asc}$ at the time
		of \ob7 to obtain a difference between the predicted and locally measured $T_{\rm asc}$
		of $\Delta T = 0.002(8)$ days. Since the predicted $T_{\rm asc}$ and the local
		measurement are consistent, and the uncertainty is smaller than half the
		orbital period to within a 95\% confidence level, we can coherently connect the
		orbital phase between these outbursts. Setting the estimated number
		of cycles between these outbursts, $N=(T_{\rm asc, 7} - T_{\rm asc, 6})/P_b$,
		to its nearest integer, then gives an accurate measurement of the orbital
		period of $P_b = 3457.892(2)$ s.

		Using the phase-connected orbital period estimate we can describe all outbursts with {a}
		single orbital model. We therefore perform a joint-fit to the data using a timing
		model in which the orbital parameters are coupled, but the frequency
		is left free per outburst. This approach gives high accuracy measurements
		of the orbital parameters, which are presented in Table \ref{tab:jointfit}.

\begin{deluxetable}{llll}
    % Preamble
    \tabletypesize{\scriptsize}
    \tablecaption{Joint-fit Timing Solution
    \label{tab:jointfit} }
    \tablewidth{1.0\linewidth}

    % Header
    \tablehead{%
	    Parameter &  Value & Statistical  & Systematic \\
	              &        & uncertainty & uncertainty 
	}

    % Data
	\startdata
        $P_b$ (s)					& 3457.8929 		& $7\E{-4}$ \\[1pt]
        $|\dot{P}_b|$ (s s$^{-1}$)	& $\leq 8\E{-11}$ 	\\[1pt]
        $A_x \sin(i)$ (lt-ms)		& 6.14 				& 0.01 \\[1pt]
        $T_{\rm asc}$ (MJD)			& 55318.04809 		& $2\E{-5}$ \\[3pt]
        $\nu_{3}$ (Hz)			& 205.892177			& $3\E{-6}$     \\[1pt]
        $\nu_{6}$ (Hz)			& 205.89217619 			& $1.1\E{-7}$	& $2\E{-6}$ 	\\[1pt]
        $\nu_{7}$ (Hz) 			& 205.8921758 			& $7\E{-7}$    	& $7\E{-6}$		\\[1pt]
        $\nu_{9}$ (Hz)			& 205.892185 			& $1.7\E{-5}$	\\[1pt]
        $\nu_{10}$ (Hz)			& 205.89208 			& $3\E{-5}$		\\[1pt]
        $\nu_{11\text{a}}$ (Hz) & 205.89212				& $4\E{-5}$		\\[1pt]
        $\nu_{11\text{b}}$ (Hz) & 205.89221				& $2\E{-5}$		\\[1pt]
        $|\dot\nu|$ (\hps{1})	& $\leq 5 \times 10^{-13}$
    \enddata

    % Postamble
    \tablecomments{%
	 	{Joint-fit timing solution with coupled orbital parameters ($\chi^2/\mbox{dof} = 41/59$).}
        $\nu_i$ gives the frequency of the $i$-th outburst, with
        $\nu_{11\text{a}}$ referring to the first observation of \ob11 and $\nu_{11\text{b}}$
        to the second (also see Table \ref{tab:outbursts}). 
    }
\end{deluxetable}

	\subsection{Spin frequency analysis}
		Because in a joint-fit approach the orbital parameters are fit to
		all data, and only the frequency is measured locally per outburst,
		the correlation between the orbit and spin parameters
		that occurs for short observations is no longer an issue. This method
		therefore allows the frequency to {be} measured in additional outbursts.
		The frequency measurements, shown in Table \ref{tab:jointfit}, are 
		consistent within their respective uncertainties, and place a 95\% confidence
		level upper limit on the frequency derivative of 
		$\abs{\dot\nu} \lesssim 5\E{-13}$ \hps{1}.
		
		As was done for the orbital phase, we may also attempt to connect
		the pulse phase between outbursts. To do this we construct a single pulse
		profile using all data of an outburst, from which we measure an averaged 
		pulse arrival time.
		Starting from \ob6, which gives the most accurate frequency measurement,
		we then propagate the timing model to predict the pulse arrival time
		for the other outbursts. In this analysis, however, there are 
		additional effects that contribute to the predicted phase or its 
		uncertainty, which need to be accounted for.
				
		The pulsations may show a frequency derivative that contributes
		to the phase difference between two outbursts. {To be conservative 
		we need to allow for the largest plausible frequency derivative, so 
		while other AMXPs with a measured long term frequency evolution appear 
		to be consistent with spin-down down due to magnetic dipole braking, we
		note that we cannot assume this is also the case for NGC 6440 X-2;
		other torquing mechanisms may be present that produce a larger 
		change of spin frequency (e.g. \citealt{dAngelo2012, Mahmoodifar2013}).
		Typically torquing mechanisms are comparable or weaker that the accretion 
		process, so we estimate the largest plausible (absolute) spin derivative}
		from  standard accretion theory \citep{Patruno2012b}
		\begin{align}
    		\label{eq:spinup}
        		\dot{\nu} &\simeq 4.2 \E{-14} \gamma_B^{1/2} 
    				\nonumber \\ &\times
    					\rbr{ \frac{ \dot{M} }{ 2 \times 10^{-10} \Msol \mbox{ yr}^{-1} } }^{6/7}
        				\rbr{ \frac{ B }{ 10^{8} \mbox{ G} } }^{2/7}  
    				\nonumber \\ &\times
    					\rbr{ \frac{ M }{ 1.4~\Msol}         }^{3/7} 
        				\rbr{ \frac{ R }{ 10 \mbox{ km} }      }^{6/7}
       				\mbox{ Hz s}^{-1}
       	\end{align}
        where $\gamma_B\simeq0.3-1$ parameterizes the uncertainty in the angular momentum at 
        the inner edge of the accretion disk \citep{Psaltis1999}, $\dot{M}$ gives 
        the mass accretion rate and $M$, $R$ and $B$ give the mass, radius and magnetic
        field strength of the neutron star, respectively.
		Assuming the pulse frequency derivative is of this order, it can
		cause a phase offset of $\sim1$ cycle over the 80 day interval between 
		\ob6 and \ob7, and so the possibility of a frequency derivative needs
		to be accounted for. 
		As the frequency derivative is the second unknown contributing to the phase
		(the frequency being the first unknown) we need to consider the phase prediction
		for two outbursts to determine both parameters. 
		
		The phase residuals of the individual outbursts are subject
		to systematic uncertainties. As demonstrated by \citet{Patruno2009b}, 
		the pulse phase shows a correlation with X-ray flux in most AMXPs. This 
		correlation can be understood as the instantaneous accretion rate causing 
		an offset in the hotspot position and thus a bias in the phase residual.
		If the flux changes linearly in time, it introduces a linear trend in the 
		phase residuals that the standard rms-minimization method corrects for by 
		adjusting the pulse frequency. This effect is particularly relevant for \ob6 
		and \ob7, which both show a decay in flux. 
		{However,} because that flux decay is nearly linear {as well} we cannot constrain the potential effect
		of a phase-flux relation from the data of NGC 6440 X-2 alone. {A strong correlation
		between phase and flux may well be present, but because it is completely degenerate with
		the pulse frequency in the timing model the phase bias cannot be distinguished from phase 
		evolution intrinsic to the pulsar.} To still obtain an approximate
		estimate for the size of this uncertainty we consider that for most AMXPs the
		phase bias due to the flux is much less than one cycle \citep{Patruno2009b}
		and adopt a maximum phase offset of $\delta\phi = 0.15$ cycles. The systematic uncertainty
		in frequency is then given as $\sigma_\nu = \delta\phi / \delta t$, where $\delta t$
		is the timespan of the considered outburst sampled by observations.
		
		Due to the phase-flux relation the pulse phase measured for an outburst consists
		of the underlying spin phase and the flux-induced bias. The phase propagation
		based on the timing model, however, only applies to the part due to the neutron
		star spin, and not the bias.
		If the phase-flux relation is indeed introduced by a geometrical effect
		such as a drifting hotspot position, then we might expect that when averaging
		over an entire outburst, the bias will average to a mean value determined by
		the system accretion geometry which is the same for all outbursts. In other
		words; if the outburst is sufficiently well sampled, then we may be able to calibrate
		the flux-induced phase bias, such that we can calculate a reference phase that is stable between
		outbursts.	
		For NGC 6440 X-2, however, the observational sampling
		of the outbursts is very poor, and consequently the measured pulse phase 
		may be offset from the outburst-long average.
		Additionally, because other AMXPs, most notably SAX J1808.4--3658, have been 
		observed to show different phase-flux relations for different outbursts 
		\citep{Patruno2009b}, the size and direction of the bias are essentially unknown
		and cannot be trivially 
		corrected for by considering the flux difference between outbursts. As such, the flux-induced
		bias on the typical pulse phase of an outburst has considerable uncertainty.
		To get at least a rough estimate of this uncertainty we adopt the same $\delta\phi = 0.15$
		cycles noted above. For \ob6 this uncertainty is roughly 10 times larger than 
		the statistical uncertainty on the averaged pulse arrival time.
								
		Taking the noted considerations into account we can propagate the pulse
		phase from \ob6 to \ob7. We then find an uncertainty on the residual phase
		of $\sigma_\phi = 14.4$. The number of trials that we need to consider is
		then calculated as
		\begin{equation}
			n = 2 ( z\sigma_\phi + \varphi_{\dot\nu} ) + 1,
		\end{equation}
		where $z$ gives the significance interval for the usual two-sided confidence
		level $\conf = \mbox{erf}\rbr{z/\sqrt{2}}$,
		$\varphi_{\dot\nu}=1$ gives the phase offset introduced by frequency
		derivative {as estimated above}, the factor two accounts for the fact that the phase offset
		can be positive and negative, and the added one accounts for the
		central trial that has a phase offset of less than half a cycle.  
		We then find that for a 95\% confidence level, we need 
		to consider 59 possible way to connect the phases between these outbursts. Similarly we 
		propagate the pulse phase from \ob6 to \ob5 to find an uncertainty of 
		$\sigma_\phi = 24.8$, implying we need to consider $100$ possible ways to connect
		those phase measurements. Hence, there are $5900$ combinations of 
		$\nu$ and $\dot\nu$ that connect the phases of the three considered outbursts.
		To test these possible solutions we can propagate each trial solution to the
		other outbursts and calculate the $\chi^2$ of the phase residuals to see which of them
		are statistically acceptable. However, due to the large phase uncertainty introduced by 
		the phase bias we find that {even for the trial solution that phase connects two outbursts} 
		the accumulated error on phase residuals at the time of the other outbursts is larger than 0.5 
		cycles even for the closest outburst. As such we cannot test the different trial solutions and must 
		conclude that all 5900 considered options are statistically acceptable.
		
		\begin{figure}[t]
			\centering
			\includegraphics[width=\linewidth]{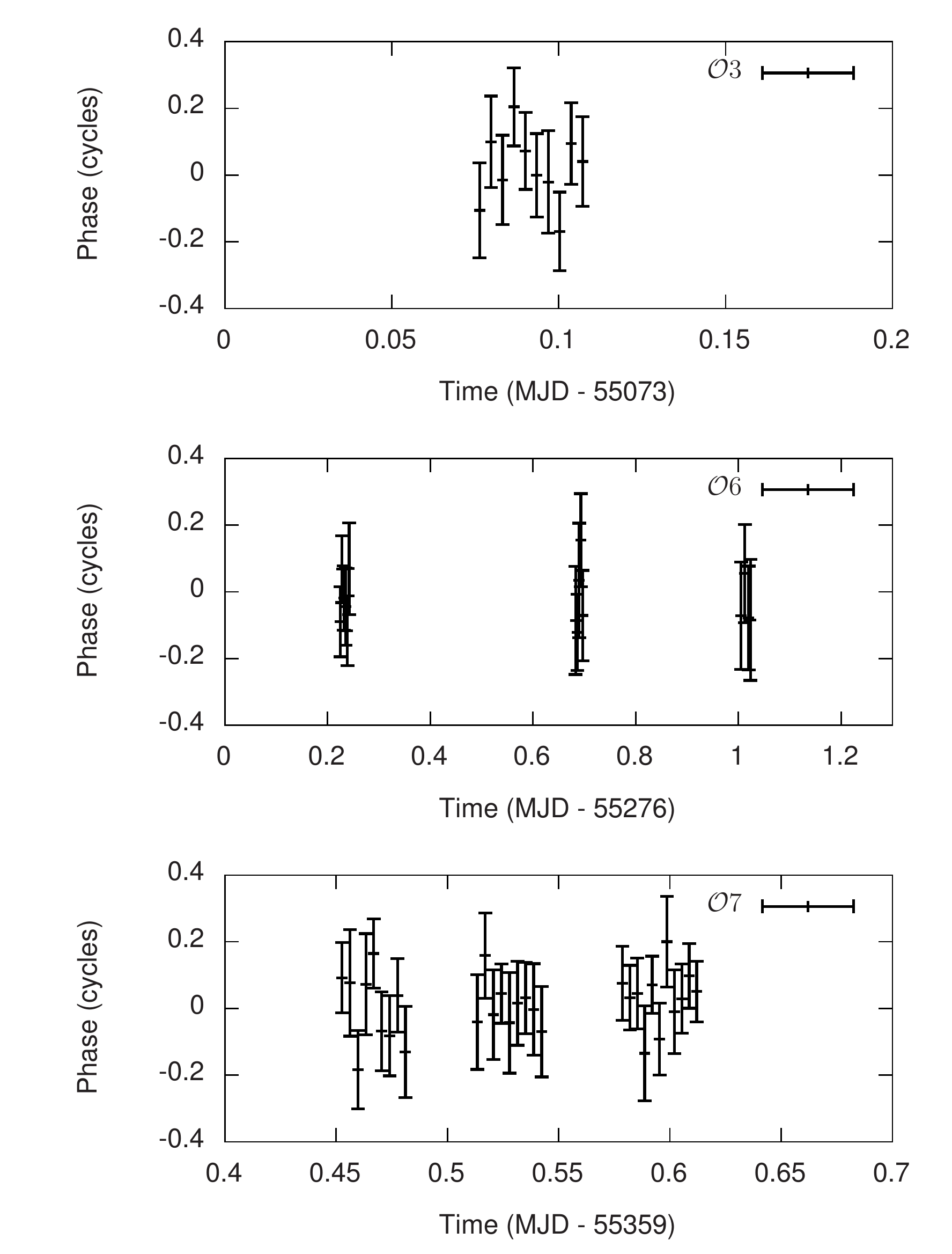}
			\caption{%
				Phase residuals of the main outbursts for the phase connected 
				timing solution assuming the the flux-induced phase bias
				is zero. \\
			}			
			\label{fig:residuals}
		\end{figure}

		It is clear that the systematic uncertainty due to the phase-flux relation
		plays an important role, and that in the case of NGC 6440 X-2, it
		becomes prohibitive in the analysis of the long term spin evolution.		
		Future observations may be able to establish the phase-flux relation in
		this source, allowing the size of the flux dependent phase bias 
		to be calibrated per outbursts, thereby reducing the systematic uncertainties or
		potentially eliminating them altogether.		
		To investigate whether a coherent pulse phase connection would then be
		feasible, we consider the scenario that the phase bias is zero in
		NGC 6440 X-2. Then, by accounting only for the statistical uncertainty on 
		the frequency measurements, the number of trials we need to consider drops 
		drastically from $100 \times 59$ to $8 \times 6 = 48$. Additionally, without
		the phase uncertainty due the flux, the error on the residual phases 
		accumulates much slower, such that the {trial} timing solutions can be coherently extrapolated
		to \ob2 and \ob10. Comparing each of the 48 solutions to the outbursts
		\ob2 through \ob10, we find that there is only one combination of $\nu$ and $\dot\nu$ 
		that provides an {statistically} acceptable fit {($\chi^2/\mbox{dof} = 66/70$)}. 
		Optimizing this solution by also refitting the orbital parameters and allowing
		for a second derivative on the pulse frequency, we find a pulse phase coherent
		timing solution that describes all data (including \ob11). We then have
			$\nu = 205.8921762261(3)$ Hz 
		with
			$\dot\nu = 1.179(3)\E{-14}$ \hps{1}
		and a second pulse frequency derivative of 
			$\ddot\nu = -3.7(1)\E{-23}$ \hps{2} 
		(see Figure \ref{fig:residuals} for the phase residuals).
		
		We stress that this result does not mean that the quoted solution {necessarily} gives 
		the {only} correct description of this system. What it shows is that if the flux-induced phase bias 
		can be calibrated, the correct phase coherent solution can be found. The solution 
		mentioned above then gives the timing solution if the phase bias can be shown to 
		be zero for this source.

\section{Discussion}
\label{sec:discussion}
	We analysed the coherent pulsations of the accreting millisecond pulsar
	NGC 6440 X-2 for all outbursts observed with {\it RXTE}. We find that in
	the ten observed outbursts, the fundamental pulse amplitude varies from 5 to 15\% rms,
	whereas the the second harmonic, if detected, has an amplitude of $\lesssim2\%$ rms.
	
	We have improved the orbital and spin parameter measurements of NGC 6440 X-2. Within 
	the uncertainty of our measurements we find no evidence of a spin derivative to an upper limit 
	of $5\E{-13}$ \hps{1}. This limit on the spin frequency derivative is larger than the 
	expected spin-up or spin-down effects in AMXPs \citep[see, e.g.][]{Patruno2012b}, and 
	therefore does not constrain the neutron star spin evolution.

	Through the coherent connection of the orbital phase, we measured the orbital
	period to high accuracy, but found no evidence of an orbital period derivative. 
	These measurements are in line with expectations from binary evolution. As 
	NGC 6440 X-2 is in an ultra-compact binary with a white dwarf companion star 
	\citep{Altamirano2010a}, its orbital evolution is expected to be dominated by 
	the loss of orbital angular momentum through the emission of gravitational 
	waves \citep{Kraft1962}, with the evolution proceeding on a timescale of 
	\citep{Paczynski1962}
	\begin{equation}
		T_{\rm GW} = 50
			\frac{ \rbr{M+M_{\rm C}}^{{1/3}} }{ M M_{\rm C} }
			P_{b}^{8/3}~\mbox{Gyr},
	\end{equation}
	where $M$ the neutron star mass, $M_{\rm C} \simeq 0.0076 \Msol$
	the companion star mass and $P_b$ is expressed in days. Assuming a 
	canonical $1.4\Msol$ neutron star we find a timescale of $1.4$ Gyr, 
	which implies an orbital derivate of $\dot{P} \sim 8\E{-14}$ \sps{1}, much smaller 
	than the upper limit we find in this work.

	The short recurrence time of NGC 6440 X-2 hints at the possibility that
	like the orbital phase, the pulse phase may also be coherently connected 
	across outbursts. Such a coherent phase connection would allow the spin 
	frequency evolution of this source to be measured to high accuracy. We 
	investigated the possibility of such a pulse phase connection, {but
	could not find a unique timing solution. The flux induced phase bias
	known to exist in most AMXPs introduces systematic uncertainties in both 
	the frequency and phase measurements, which need to be taken into account. 
	Unfortunately, due to the sparse observational sampling of NGC 6440 X-2's 
	outbursts, the phase bias is degenerate with the pulse frequency
	parameter in the timing model. Consequently the presence and size of this 
	phase bias could not be constrained.}
	
    Although we could not exclude the presence of a flux induced phase bias,
    we did attempt to phase connect the pulsations under the assumption that
    this effect does not occur for NGC 6440 X-2. We then find that a phase 
    connection is possible and obtain a timing solution for a long term 
    \textit{spin-up} of the pulsar.
    Assuming typical outbursts last for about four days, the average 
    duty cycle of NGC 6440 X-2 is about 5\%, although we note this number
    may be slightly larger, as we may not have observed all outbursts. For this
    duty cycle the accretion spin-up during outburst would have to be about
    $2.4\E{-13}$ \hps{1}. Based on accretion theory (eq. \ref{eq:spinup}) such
    a large spin-up might be possible, but requires significant fine-tuning
    of the neutron star parameters (large neutron star mass and magnetic field 
    strength). So, while this long term spin-up is possible, it seems to us that 
    it is more likely that it indicates the underlying assumption of zero 
    phase bias does not hold and flux induced phase uncertainties indeed affect
    the observations of this AMXP.
    
	Our analysis {highlights} the important role played by the systematic effects
	in pulse phase due to the X-ray flux when considering coherent timing of different
	outbursts. This applies not just to NGC 6440 X-2, but to any accreting 
	millisecond pulsar.
	
	We {suggest} that if future observations {provide a better sampling of NGC
	6440 X-2's outbursts} and are able to {constrain the size of} the phase bias a 
	coherent phase connection {may} be possible, warranting a closer investigation of 
	this interaction between pulse phase and instantaneous X-ray flux and the mechanism 
	by which it arises.

\acknowledgments
PB and MvdK acknowledge support from the Netherlands Organisation for 
Scientific Research (NWO). AP is supported by a NWO Vidi Fellowship.

\bibliographystyle{apj}

\begin{thebibliography}{24}
\expandafter\ifx\csname natexlab\endcsname\relax\def\natexlab#1{#1}\fi

\bibitem[{{Altamirano} {et~al.}(2010{\natexlab{a}}){Altamirano}, {Patruno},
  {Heinke}, {Linares}, {Markwardt}, \& {Strohmayer}}]{Altamirano2010b}
{Altamirano}, D., {Patruno}, A., {Heinke}, C., {et~al.} 2010{\natexlab{a}}, The
  Astronomer's Telegram, 2500, 1

\bibitem[{{Altamirano} {et~al.}(2009){Altamirano}, {Strohmayer}, {Heinke},
  {Markwardt}, {Swank}, {Pereira}, {Smith}, {Wijnands}, {Linares}, {Patruno},
  {Casella}, \& {van der Klis}}]{Altamirano2009}
{Altamirano}, D., {Strohmayer}, T.~E., {Heinke}, C.~O., {et~al.} 2009, The
  Astronomer's Telegram, 2182, 1

\bibitem[{{Altamirano} {et~al.}(2010{\natexlab{b}}){Altamirano}, {Patruno},
  {Heinke}, {Markwardt}, {Strohmayer}, {Linares}, {Wijnands}, {van der Klis},
  \& {Swank}}]{Altamirano2010a}
{Altamirano}, D., {Patruno}, A., {Heinke}, C.~O., {et~al.} 2010{\natexlab{b}},
  \apjl, 712, L58

\bibitem[{{Bildsten}(1998)}]{Bildsten1998b}
{Bildsten}, L. 1998, \apjl, 501, L89

\bibitem[{{Bildsten}(2002)}]{Bildsten2002}
---. 2002, \apjl, 577, L27

\bibitem[{{D'Angelo} \& {Spruit}(2012)}]{dAngelo2012}
{D'Angelo}, C.~R., \& {Spruit}, H.~C. 2012, \mnras, 420, 416

\bibitem[{{Haskell} \& {Patruno}(2011)}]{Haskell2011}
{Haskell}, B., \& {Patruno}, A. 2011, \apjl, 738, L14

\bibitem[{{Heinke} {et~al.}(2009){Heinke}, {Jonker}, {Wijnands}, {Deloye}, \&
  {Taam}}]{Heinke2009}
{Heinke}, C.~O., {Jonker}, P.~G., {Wijnands}, R., {Deloye}, C.~J., \& {Taam},
  R.~E. 2009, \apj, 691, 1035

\bibitem[{{Heinke} {et~al.}(2010){Heinke}, {Altamirano}, {Cohn}, {Lugger},
  {Budac}, {Servillat}, {Linares}, {Strohmayer}, {Markwardt}, {Wijnands},
  {Swank}, {Knigge}, {Bailyn}, \& {Grindlay}}]{Heinke2010}
{Heinke}, C.~O., {Altamirano}, D., {Cohn}, H.~N., {et~al.} 2010, \apj, 714, 894

\bibitem[{{Kraft} {et~al.}(1962){Kraft}, {Mathews}, \&
  {Greenstein}}]{Kraft1962}
{Kraft}, R.~P., {Mathews}, J., \& {Greenstein}, J.~L. 1962, \apj, 136, 312

\bibitem[{{Leahy} {et~al.}(2008){Leahy}, {Morsink}, \& {Cadeau}}]{Leahy2008}
{Leahy}, D.~A., {Morsink}, S.~M., \& {Cadeau}, C. 2008, \apj, 672, 1119

\bibitem[{{Mahmoodifar} \& {Strohmayer}(2013)}]{Mahmoodifar2013}
{Mahmoodifar}, S., \& {Strohmayer}, T. 2013, \apj, 773, 140

\bibitem[{{Nelson} \& {Rappaport}(2003)}]{Nelson2003}
{Nelson}, L.~A., \& {Rappaport}, S. 2003, \apj, 598, 431

\bibitem[{{Paczy{\'n}ski}(1967)}]{Paczynski1962}
{Paczy{\'n}ski}, B. 1967, Acta Astron., 17, 287

\bibitem[{{Patruno} {et~al.}(2010){Patruno}, {Altamirano}, \&
  {Messenger}}]{Patruno2010c}
{Patruno}, A., {Altamirano}, D., \& {Messenger}, C. 2010, \mnras, 403, 1426

\bibitem[{{Patruno} {et~al.}(2012){Patruno}, {Bult}, {Gopakumar}, {Hartman},
  {Wijnands}, {van der Klis}, \& {Chakrabarty}}]{Patruno2012a}
{Patruno}, A., {Bult}, P., {Gopakumar}, A., {et~al.} 2012, \apjl, 746, L27

\bibitem[{{Patruno} \& {D'Angelo}(2013)}]{Patruno2013}
{Patruno}, A., \& {D'Angelo}, C. 2013, \apj, 771, 94

\bibitem[{{Patruno} \& {Watts}(2012)}]{Patruno2012b}
{Patruno}, A., \& {Watts}, A.~L. 2012, in Timing neutron stars: pulsations,
  oscillations and explosions, ed. T.~{Belloni}, M.~{Mendez}, \& C.~M. {Zhang}
  (in press)

\bibitem[{{Patruno} {et~al.}(2009){Patruno}, {Wijnands}, \& {van der
  Klis}}]{Patruno2009b}
{Patruno}, A., {Wijnands}, R., \& {van der Klis}, M. 2009, \apjl, 698, L60

\bibitem[{{Poutanen} \& {Gierli{\'n}ski}(2003)}]{Poutanen2003}
{Poutanen}, J., \& {Gierli{\'n}ski}, M. 2003, \mnras, 343, 1301

\bibitem[{{Psaltis} {et~al.}(1999){Psaltis}, {Belloni}, \& {van der
  Klis}}]{Psaltis1999}
{Psaltis}, D., {Belloni}, T., \& {van der Klis}, M. 1999, \apj, 520, 262

\bibitem[{{Psaltis} {et~al.}(2014){Psaltis}, {{\"O}zel}, \&
  {Chakrabarty}}]{Psaltis2014a}
{Psaltis}, D., {{\"O}zel}, F., \& {Chakrabarty}, D. 2014, \apj, 787, 136

\bibitem[{{Rots} {et~al.}(2004){Rots}, {Jahoda}, \& {Lyne}}]{Rots2004}
{Rots}, A.~H., {Jahoda}, K., \& {Lyne}, A.~G. 2004, \apjl, 605, L129

\bibitem[{{van Straaten} {et~al.}(2003){van Straaten}, {van der Klis}, \&
  {M{\'e}ndez}}]{Straaten2003}
{van Straaten}, S., {van der Klis}, M., \& {M{\'e}ndez}, M. 2003, \apj, 596,
  1155

\end{thebibliography}

\end{document}